\documentclass[11pt,twoside]{article}
\usepackage{graphicx,epsfig,natbib,epstopdf}
\usepackage{CS18}
\begin{document}
%
%
% NTA: Here is an example JPG illustration.  This only works under PDFLaTeX, not straight LaTeX.
%

\title{Magnetic field on the brown dwarf LSR J18353790+3259545}
\author{O. Kuzmychov$^1$, S. Berdyugina$^{1,2}$ \& D. Harrington$^{1,3}$}
\vspace{0.2cm}
\affil{$^1$ Kiepenheuer-Institut f\"ur Sonnenphysik, Freiburg, Germany}
\affil{$^2$ NASA Astrobiology Institute, University of Hawaii, USA}
\affil{$^3$ Institute for Astronomy, University of Hawaii, USA}

\abstract{
We model the full-Stokes spectrum of the brown dwarf LSR J18353790+3259545 in the bands of the diatomic molecules CrH, TiO, and FeH in order to infer its magnetic properties. The models are then compared to the observational data obtained with the Low Resolution Imaging Spectrograph operated in the spectropolarimetric mode (LRISp) at Keck observatory. Our preliminary analysis shows that the brown dwarf considered possesses a magnetic field of the order of $2-3$~kG.

%For doing this, we model polarimetric spectra of diatomic molecules - CrH, FeH, and TiO - and atomic lines. The objects observed exhibit transient but periodic radio pulses that are possibly driven by electron-cyclotron maser (Hallinan et al. 2007, 2008). For this mechanism to work, a few kG magnetic field is however required. In order to examine whether these objects possess such a strong magnetic field, we employ molecular spectropolarimetry (Berdyugina et al. 2000, 2005; Afram et al. 2008; Kuzmychov \& Berdyugina 2013), that allows us to explore the magnetic fields in cool atmospheres of brown dwarfs and exoplanets. We are able to constrain the magnetic field strengths in these objects.
}

\section{Introduction}

Radio observations by \citet{berger2002} and \citet{berger2006} revealed a number of brown dwarfs which show persistent radio emission and flares, providing an indication for their magnetic activity. Though several mechanisms can drive radio emission in these objects, the electron cyclotron maser instability is the most likely one \citep{hallinan2008}. For this mechanism to work a strong magnetic field of the order of $3$~kG is required.

In order to examine the magnetosphere of the radio pulsating brown dwarf LSR J18353790+3259545 (hereafter, LSR J1835), we employ our spectropolarimetric expertise developed for the CrH molecule \citep{kuzmychov2013}. We calculate a synthetic full-Stokes spectrum of the (0,0) vibrational band of the CrH molecule in the presence of a magnetic field, and we then compare it to the observational data. In addition to the lines from the (0,0) band of the CrH molecule, we include the TiO and FeH lines and the atomic lines that blend with the CrH (0,0) band. 

The full-Stokes spectrum of LSR J1835 was obtained with the Low Resolution Imaging Spectrograph in the polarimetric mode (LRISp) at Keck observatory in August 2012 at three different rotational phases. The process of data acquisition and data reduction is described briefly in \citet{kuzmychov2013a} and it will be described in details in a forthcoming paper by Harrington et al. Only one rotational phase is analyzed here.

\section{Modeling the data}
We use the code STOPRO (described by \citet{solanki1987, frutiger1999, berdyugina2003}) to perform the radiative transfer calculations for the spectral lines of the diatomic molecules in the presence of a magnetic field. Our line list includes about $500$ lines from the (0,0) band of the CrH, $400$ TiO lines, $40$ lines from the (1,0) band of the FeH, and about $60$ strong atomic lines, all lines falling into the wavelength region $8600-8700$~\AA. For our radiative transfer calculations we used the BT-Settl atmospheric model grid by \citet{allard2010}. 

\begin{figure}[ht!]
\centering
%
% NTA: Please update photo file name here.  Use a unique identifier.
%
\includegraphics[width=\textwidth]{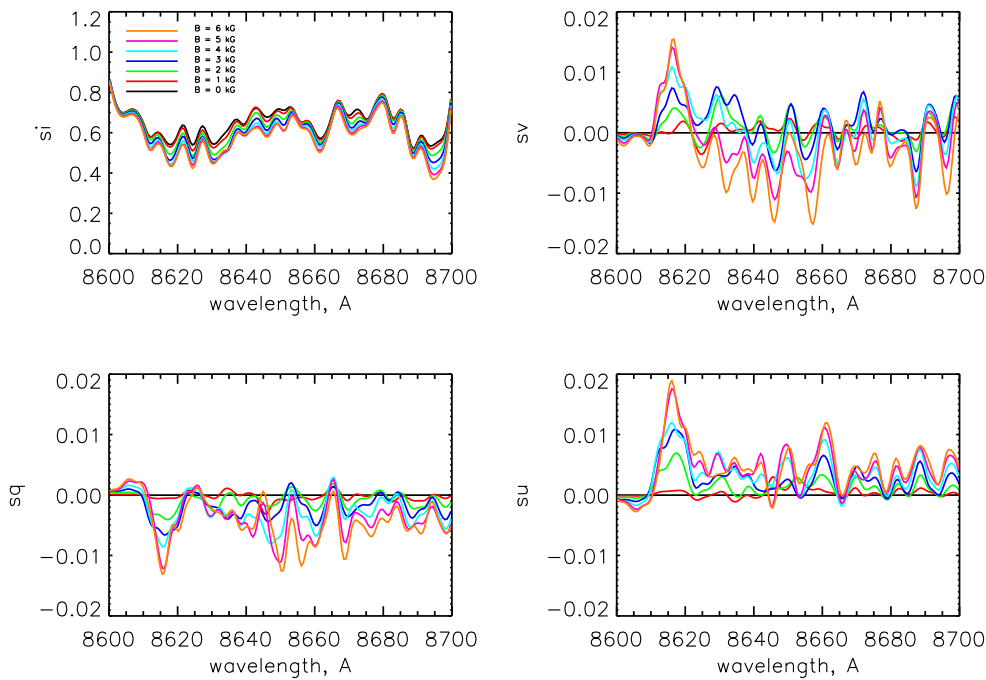}
%
% NTA: Please update caption here.  Identify any people you know.  Please provide institution.
%
\caption{Synthetic Stokes profiles $I/I_c$ (upper left), $V/I_c$ (upper right), $Q/I_c$ (lower left), and $U/I_c$ (lower right) calculated for different magnetic field strengths.}
\label{fig: fig1}
\end{figure}

Figure~\ref{fig: fig1} shows the synthetic Stokes profiles calculated for different magnetic field strengths. The inclination of the magnetic field is set to $112.5^\circ$ and its azimuth to $67.5^\circ$. The atmospheric model used was calculated for $T_\textrm{eff}=2500$~K, $\log g=5.0$ and solar abundances of the chemical elements. An instrumental broadening of 3~\AA, which corresponds roughly to that of LRISp, and $v\sin i=40$~kms$^{-1}$ were taken into account.

While the intensity signal Stokes $I/I_c$ (upper left, Fig.~\ref{fig: fig1}) does not vary much with the strength of a magnetic field, the polarimetric signals Stokes $V/I_c$, $Q/I_c$, and $U/I_c$ does vary a lot: From no signal in the absence of a magnetic field up to 2 percent at $6$~kG. Furthermore, the polarimetric signals produced by the lines of the diatomic molecules considered show a unique asymmetric shape for a given magnetic field strength. This fact allows us an unambiguous estimation of the magnetic field strength when the models are compared to the observational data\citep{kuzmychov2013}.

\section{Results}
We present an estimate of the magnetic field strength for LSR J1835 based on our preliminary analysis of the Stokes profiles observed at one rotational phase. 

\begin{figure}[ht!]
\centering
\includegraphics[width=\textwidth]{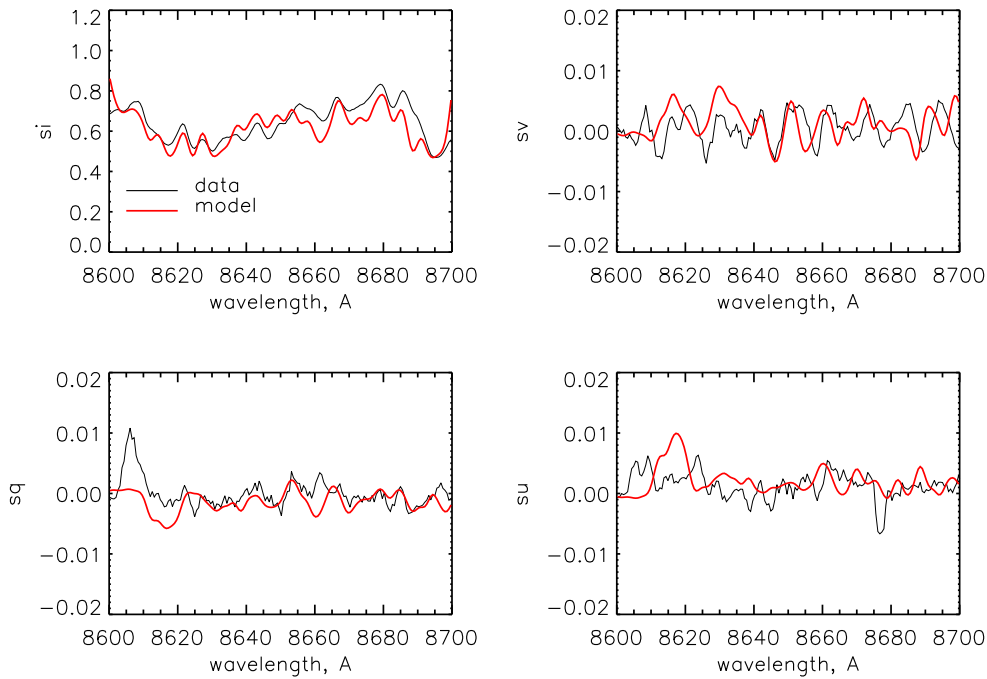}
\caption{Comparison of our model (red line) to the full Stokes spectrum of LSR J1835 (black line).}
\label{fig: fig2}
\end{figure}

Figure~\ref{fig: fig2} shows a comparison of our model to the observational data. The model is calculated for the effective temperature $T_\textrm{eff}=2500$~K and surface gravity $\log g=5.0$. The strength and orientation of the magnetic field ($B=2.5$~kG, inclination of $B$ $112.5^\circ$, and azimuth of $B$ $67.5^\circ$) were chosen to match the strength and shape of the polarimetric signal observed.  The instrumental broadening of $3$~\AA, which we derived from the data reduction procedure, and rotational velocity $v\sin i=40$~kms$^{-1}$ \citep{deshpande2012} were taken into account.

Based on the analysis outlined, we preliminarily estimate the magnetic field strength of LSR J1835 to $2.5$~kG. This value is in a good agreement with that obtained from the previous observations of the radio pulsating brown dwarfs \citep{berger2006, hallinan2006, hallinan2008}.

\section{Discussion}
We are going to explore the free parameter space of our models and, based on a $\chi^2$ minimization, constrain the value of the magnetic field strength and provide its error. Moreover, we intend to fully analyze the delivered instrumental performance and account for instrumental errors. 

The large spikes in the polarimetric signals observed (cf.~Fig.~\ref{fig: fig2}, Stokes $Q/I_c$ and $U/I_c$) can be attributed to non-shot-noise sources, such as cosmic ray hits. We are improving our data reduction pipeline, and we will be able to filter the noise sources more effectively in the future. 

An analysis of the three rotational phases will allow us to determine the orientation of the magnetic field in each phase and, thus, will allow us to track the magnetic region as the dwarf rotates. The results will be published in a forthcoming paper by Kuzmychov et al. 

\bibliographystyle{aa} %style aa.bst
%\citet{kuzmychov2013a} cite reference in text
%\citep{kuzmychov2013b} cite reference in braces
\bibliography{lit_list_kuzmychov} %references in crhpaper.bib

\end{document}